\documentclass[conference]{IEEEtran}
\IEEEoverridecommandlockouts
\usepackage{cite}
\usepackage{amsmath,amssymb,amsfonts}
\usepackage{algorithmic}
\usepackage{graphicx}
\usepackage{textcomp}
\usepackage{xcolor}
\usepackage{academicons}
\usepackage{overpic}
\newcommand{\orcid}[1]{\href{https://orcid.org/#1}{\textcolor[HTML]{A6CE39}{\aiOrcid}}}
\def\BibTeX{{\rm B\kern-.05em{\sc i\kern-.025em b}\kern-.08em
    T\kern-.1667em\lower.7ex\hbox{E}\kern-.125emX}}
\begin{document}

\title{StateFormer: A Multivariate Transformer for Learning History-Dependent Battery State Dynamics and Long-Horizon Health Forecasting\\ 
\thanks{This work was supported by the U.S. Department of Energy, Office of Electricity under Award Number DE-AC02-05CH11231.}
}


\author{
  \IEEEauthorblockN{
    Zhe Bai\textsuperscript{a},
    Stephen Harris\textsuperscript{b}
  }
  \IEEEauthorblockA{
    \textsuperscript{a}\emph{Applied Mathematics and Computational Research}\\
    \textsuperscript{b}\emph{Energy Storage \& Distribution}\\
    \emph{Lawrence Berkeley National Laboratory} \\
    Berkeley, CA 94720, USA\\
   \{zhebai, sjharris\}@lbl.gov}
  }


\maketitle

\begin{abstract}
This paper introduces a novel multivariate Transformer \emph{StateFormer} that forecasts degradation dynamics of large-scale battery systems. The model learns across time scales, from short-term thermal fluctuations to long-term aging trajectories, enabling accurate prediction of battery states including state of charge (SOC), state of health (SOH), or battery temperature. By capturing long-range dependencies and identifying the operating conditions that drive future degradation, the model simultaneously represents fast electrochemical and thermal processes and slow aging mechanisms within a unified framework. 
\emph{StateFormer} achieves robust and accurate predictions of battery state estimation across both synthetic and real-world datasets. It maintains high predictive performance under additive current/voltage noise levels ranging from $1\%$ to $10\%$ and a wide range of ambient temperatures in a synthetic battery fleet accommodating different types of electrode chemistry and manufacturing, as well as five years of field data collected from residential utility battery systems.
The resulting model bridges the laboratory-to-field gap and provides predictive intelligence for maintenance planning, operational optimization, and economic decision-making.
\end{abstract}


\section{Introduction}
The rapid integration of variable energy and surging power demands of future grid infrastructures rely heavily on the deployment of large-scale Battery Energy Storage Systems (BESS).
Understanding and accurately forecasting the long-term degradation of these battery fleets is a formidable while essential challenge. 
Grid-scale battery degradation unfolds over multi-year timescales under deeply coupled electrochemical, thermal, and operational feedback. Reliable monitoring and long-term forecasting of battery states, including the state of charge (SOC) and state of health (SOH), are essential for ensuring system reliability, optimizing maintenance, and extending asset lifetime. However, accurate battery state forecasting remains challenging due to the nonlinear interactions among electrochemical, thermal, and operational processes, which evolve over years under diverse conditions.

Existing data-driven approaches~\cite{how2019state, wang2021review,severson2019data} are constrained by limited training data. Laboratory datasets collected under controlled cycling fail to capture the heterogeneous operating and environmental conditions of real-world battery systems, while field data are often proprietary, incomplete, and lack ground-truth degradation labels. Moreover, most existing methods rely on representative battery models that assume homogeneous degradation across an entire fleet. In practice, battery fleets exhibit significant variability due to manufacturing tolerances, thermal gradients, and diverse operating conditions. Consequently, accurate SOH forecasting must capture the distribution of degradation trajectories across battery fleets to enable reliable warranty assessment, predictive maintenance, and fleet-scale reliability management.
To optimize future grid and data center storage, operators must reliably forecast how dispatch policies interact with site-specific environmental controls. 

These challenges motivate the development of robust AI models capable of learning long-term battery dynamics from both synthetic and field data while generalizing across diverse operating conditions and measurement uncertainties. To enable reliable deployment, such models \emph{must} remain resilient to measurement noise, abrupt operational transitions, and discontinuous system behaviors while maintaining accurate long-horizon predictions. Physics-grounded synthetic simulation platforms complement limited field measurements by generating degradation behaviors that are difficult to observe experimentally, including accelerated aging under extreme operating environment, such as high temperatures.

State-space models (SSMs)~\cite{bai2020dynamic,zhou2023deep} provide a principled framework for modeling dynamical systems by separating latent system evolution from noisy observations, while Transformer architectures~\cite{vaswani2017attention,bai2026predicting} have demonstrated remarkable capability in capturing long-range temporal dependencies. 
Building on these complementary strengths, we propose \emph{StateFormer}, a state-space transformer for long-horizon battery state forecasting. \emph{StateFormer} is validated using both a physics-grounded synthetic battery fleet spanning a wide range of ambient temperatures and multi-year field measurements from a residential battery energy storage system~\cite{figgener2024multi}. The results demonstrate accurate and robust long-horizon forecasting, strong extrapolation to unseen temperature conditions, resilience to measurement noise, and reliable generalization from simulation to real-world operation, making \emph{StateFormer} a practical framework for battery fleet monitoring and predictive maintenance.

\begin{figure*}[tbp]
\centering
\includegraphics[width=1.95\columnwidth]{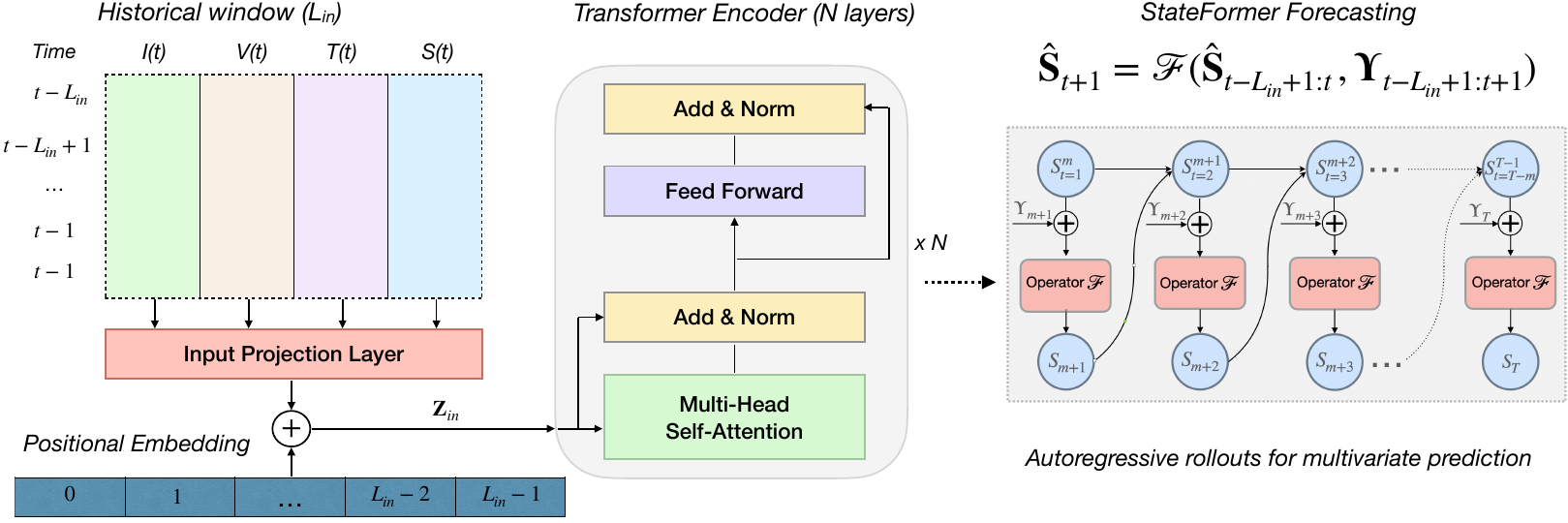}
\vspace{-.05in}
\caption{Schematic of the proposed \emph{StateFormer} framework. A sliding historical window of battery operating states is encoded into attention-based latent embeddings that capture long-term degradation dynamics. Incorporating the historical operating states, the current operating inputs with the latest battery state, the model autoregressively rolls out the battery states at the following time step.}
\label{fig1}
\vspace{-.15in}
\end{figure*}
\section{Neural Degradation Model}
A key component of the proposed framework is the neural degradation model, which captures the long-term evolution of battery health under realistic operating conditions. 
\emph{StateFormer} exploits both state-space representations and long-range attention mechanisms to model the coupled electrochemical and thermal dynamics governing battery aging. This data-driven formulation enables accurate long-horizon state (e.g., SOH) forecasting while remaining robust to measurement noise and heterogeneous operating conditions.

Fig.~\ref{fig1} illustrates the neural architecture of the proposed \emph{StateFormer} for learning long-range battery state dynamics. The battery state at time $\mathbf{S}_{t+1}$ is determined not only by its current state $\mathbf{S}_{t}$, but also by the cumulative operating states encoded in the historical window $L_{in}$ and the latest actuation or operating condition $\mathbf{\Upsilon}_{t+1}$. 
Through self-attention $\alpha_i = \mathrm{softmax}(q_{t+1}k_i^{\top})$, the model dynamically identifies the most relevant historical degradation patterns, while positional encoding preserves the temporal ordering of these states. The resulting latent representation $\mathbf{Z}_{in}$ integrates long-term degradation memory with the immediate operating input, enabling an accurate battery state estimation. Specifically,
the model learns a state transition function of the form in which the transition depends jointly on the degradation history and the current operating input as~\eqref{eq1}: 
\begin{align}
{\hat{\mathbf{S}}_{t+1}}=\mathcal{F}(\hat{\mathbf{S}}_{t-L_{in}+1:t}, \mathbf{\Upsilon}_{t-L_{in}+1:t+1}).
\label{eq1}
\end{align}
Here, $\mathbf{S}$ represents multi-variable output, such as the state-of-charge (SOC) and state-of-health (SOH) variables, and $\mathbf{\Upsilon}_t$ consists of the instantaneous operating variables, including measured current and ambient temperature. The historical SOC and SOH sequence captures the cumulative effects of irreversible degradation mechanisms, while the current electrical and thermal states characterize the immediate operating stress experienced by the battery. Combining these complementary sources of feature information, the neural model learns the nonlinear degradation dynamics directly from data without requiring explicit electrochemical or empirical aging equations. Implemented as an autoregressive Transformer, \emph{StateFormer} recursively propagates the battery health state forward in time by conditioning each prediction on previously estimated multivariate states along with the current operating inputs, thereby providing a data-driven approximation to the underlying degradation dynamics. This state-space formulation naturally captures the path-dependent nature of battery aging, enabling accurate long-horizon state forecasting under varying operating conditions and serving as the foundation for degradation-aware battery operation and optimization.

To rigorously assess long-horizon forecasting, all predictions are performed in an autoregressive rollout without state correction from future measurements. For the synthetic battery fleet, only the first $50$ hours of complete battery states are provided to initialize the model. Thereafter, \emph{StateFormer} recursively predicts SOC and SOH for the remainder of the three-year evaluation period using its own previous predictions conditioned on the measured hourly current and ambient temperature as inputs. Similarly, for the field measurements, only the first $120$ minutes of battery voltage and temperature are used for initialization. The model then recursively forecasts battery voltage and temperature over the remaining one-year period using the measured current and room temperature at each one-minute interval. This protocol evaluates the model's ability to maintain stable and accurate long-horizon forecasts without access to future ground-truth battery states.

\section{Case study}
\subsection{Synthetic fleet with inherent heterogeneity and noise}
First, we evaluate the proposed framework using a synthetic dataset generated from hour-resolved, multi-decade simulations that couple stochastic environmental conditions, market-driven dispatch, electrothermal dynamics, and fleet heterogeneity. The dataset assumes a fixed-duration, constant-power discharge strategy representative of grid-scale battery energy storage systems participating in energy arbitrage and resource adequacy services, with each battery executing at most one price-driven discharge cycle per day. We simulate the time development of SOC given by~(\ref{eq:SOC}),
\begin{align}\label{eq:SOC}
\mathrm{SOC}(t+\Delta t)=\mathrm{SOC}(t)- \frac{P_{\mathrm{grid}}(t)\Delta t}{\eta_{\mathrm{dis}}(t)E_{\mathrm{cap}}(t)},
\end{align}
where $P_{\mathrm{batt}(t)}=P_{\mathrm{grid}(t)}/\eta_{\mathrm{dis}(t)}$.
During non-dispatch hours $P_{\mathrm{grid}(t)}=0$, and therefore $P_{\mathrm{batt}}(t)=0$, so SOC remains unchanged. 
As state of health declines, the allowable SOC operating window contracts toward end-of-life limits
\begin{align}\label{eqSOCmax}
\mathrm{SOC_{max}}(t) &= \mathrm{SOC_{max,BOL}}-(\mathrm{SOC_{max,BOL}}\nonumber\\&-\mathrm{SOC_{max,EOL}})\frac{1-\mathrm{SOH}(t)}{1-\mathrm{SOH_{EOL}}},
\end{align}
and the usable SOC range shrinks as the battery ages. The increasing internal resistance raises the current required to meet a fixed power request and increases polarization losses (\emph{IR} drop), so voltage limits are encountered earlier within the nominal SOC range. Under baseline parameters, the allowable SOC window narrows from $0.05\text{-}0.95$ at the beginning of life (BOL) to $0.20\text{-}0.80$ at end of life (EOL)~\cite{rahman2024exploring,kumtepeli2024depreciation}.
The ambient temperature ${T_{amb}}$ represents the air temperature inside the battery container regulated by HVAC, and is modeled as 
\begin{align}
{T_{amb}}(t)={T_{setpoint}}+\alpha({T_{out}}(t)-{\bar{T}_{out}})+\epsilon(t)),
\end{align}
where ${T_{setpoint}}$ is the HVAC target ($22^{\circ}\mathrm{C}$ baseline),
$\alpha$ is the attenuation factor representing the fraction of outdoor temperature deviation that penetrates the container, and $\epsilon(t)$ is small-amplitude noise capturing HVAC control imprecision. 
Cell temperature determines degradation rates through Arrhenius kinetics, employing a quasi-steady-state thermal model,
\begin{align}
{T_{cell}}(t)&={T_{amb}}(t)+\Delta {T_{static}} + \Delta {T_{oper}}(t), \\
\Delta {T_{static}}&=g\cdot \Delta {T_{gradient}},
\end{align}
where $g$ represents vertical position within the rack ($0=$ bottom, $1=$ top) and ${T_{gradient}}$ is the total temperature difference from bottom to top layer. 
The operating heat generated from discharge is modeled as
\begin{align}
\label{eq8}
\Delta{T_{oper}}(t)&=K_T\cdot{P_{heat}}(t), \\
\label{eq9}{P_{heat}}(t)&={P_{grid}}(t)(\frac{1}{\eta_{dis}(t)}-1),
\end{align}
where~\eqref{eq8} relates heat generation rate to temperature rise through $K_T$, an effective steady-state thermal gain ($^{\circ}\mathrm{C}/KW$) that aggregates enclosure heat rejection under quasi-steady operating conditions. Equation~\eqref{eq9} calculates the heat generation rate from the difference between power drawn from the battery and power delivered to the grid, where the feedback loop of heat generation inside cells accelerates Arrhenius-driven degradation and further reduces its efficiency.
The total capacity loss is the sum of calendar and cycle contributions accumulated over the simulation horizon:
\begin{align}
\mathrm{SOH}(t) = 1-{Q_{loss,cal}}-{Q_{loss,cyc}}(t).
\end{align}
Calendar aging is modeled as continuous capacity loss driven by temperature and state of charge, accumulating over elapsed time. The instantaneous calendar-aging rate follows a power-law dependence modulated by temperature and SOC stress factors:
\begin{align}
\frac{d{Q_{cal}}}{dt} &=k_{\mathrm{cal}} \beta t^{\beta-1} {f_T}(T_{cell}) \mathrm{f_{SOC}}(\mathrm{SOC}),\\
f_T(T_{cell})&=\exp[\frac{{E_{a,cal}}}{R}(\frac{1}{T_{ref}}-\frac{1}{T_{cell}})],\\
\mathrm{f_{SOC}}(\mathrm{SOC})&=\exp[\alpha_{cal}\cdot( \mathrm{SOC}- \mathrm{{SOC}_{ref}})]
\end{align}
where ${Q_{cal}}$ is fractional capacity loss, $k_{cal}$ is a rate constant, and ${f_{T}},{f_\mathrm{SOC}}$ represent temperature- and SOC- dependent acceleration factors.  
The time exponent $\beta$ is treated as a configurable parameter, allowing exploration of different calendar-aging behaviors under grid-relevant operating conditions. Calendar aging accumulates continuously, including during idle periods, with the rate determined by the prevailing thermal and SOC conditions. Cycle aging accumulates proportionally to energy throughput with Arrhenius temperature acceleration:
\begin{align}
\frac{d{Q_{cyc}}}{dt} &=k_{\mathrm{cyc}} \frac{\|{P_{batt}}(t)\|}{\mathrm{E_{cap}}} {f_{T,cyc}},\\
{f_{T,cyc}}&=\exp[\frac{{E_{a,cyc}}}{R}(\frac{1}{T_{ref}}-\frac{1}{T_{cell}})],
\end{align}
where cycling damage is coupled to instantaneous cell temperature, capturing the physical reality that cycling at elevated temperatures accelerates degradation. 
Discharge efficiency degrades linearly with the SOH:
\begin{align}
\mathrm{\eta_{dis}}(t) &= \mathrm{\eta_{dis,BOL}}-(\mathrm{\eta_{dis,BOL}}-\mathrm{\eta_{dis,EOL}})\frac{1-\mathrm{SOH}(t)}{1-\mathrm{SOH_{EOL}}}.
\end{align}
This coupled mechanism creates an electro-thermal feedback loop as internal resistance grows with aging, discharge efficiency declines, increasing resistive heat generation and thereby accelerating Arrhenius-driven degradation~\cite{koller2013defining}.
Assets are retired when state of health falls below a fixed, configurable threshold~\cite{liu2025warranties}, $\mathrm{SOH}(t)\leq\mathrm{SOH_{EOL}}$. The baseline threshold is set to $\mathrm{SOH_{EOL}}=0.7$, rendering grid-scale systems under typical market and safety constraints.

In the fleet simulation, we generate populations of assets sharing identical market environments and dispatch schedules but differing in intrinsic characteristics. Each asset receives a quality factor sampled from $q_\mathrm{factor}\sim\mathcal{N}(1.0,\sigma_{\mathrm{qual}})$. Degradation rate constants are divided by this factor, representing cell-to-cell variability in effective reaction kinetics, interfacial stability, and defect density reflecting manufacturing tolerances for commercial BMS measurement uncertainty~\cite{fasolato2025analyzing,shete2021battery,che2025diagnostic}.

The synthetic dataset consists of 50 unique battery cells, each simulated under five ambient temperatures ($25, 30, 35, 40$, and $45^{\circ}\mathrm{C}$). To evaluate the robustness of \emph{StateFormer}, additive Gaussian noise with levels ranging from $1\%$ to $5\%$ was independently applied to the measured current and voltage signals. To assess the model's extrapolation capability, \emph{StateFormer} was trained using data from the lower-temperature conditions ($25\text{–}35^{\circ}\mathrm{C}$) and evaluated on unseen operating conditions at elevated temperatures of $40$ and $45^{\circ}\mathrm{C}$.

\begin{figure}[tbp]
\centering
\begin{overpic}[width=0.95\columnwidth]{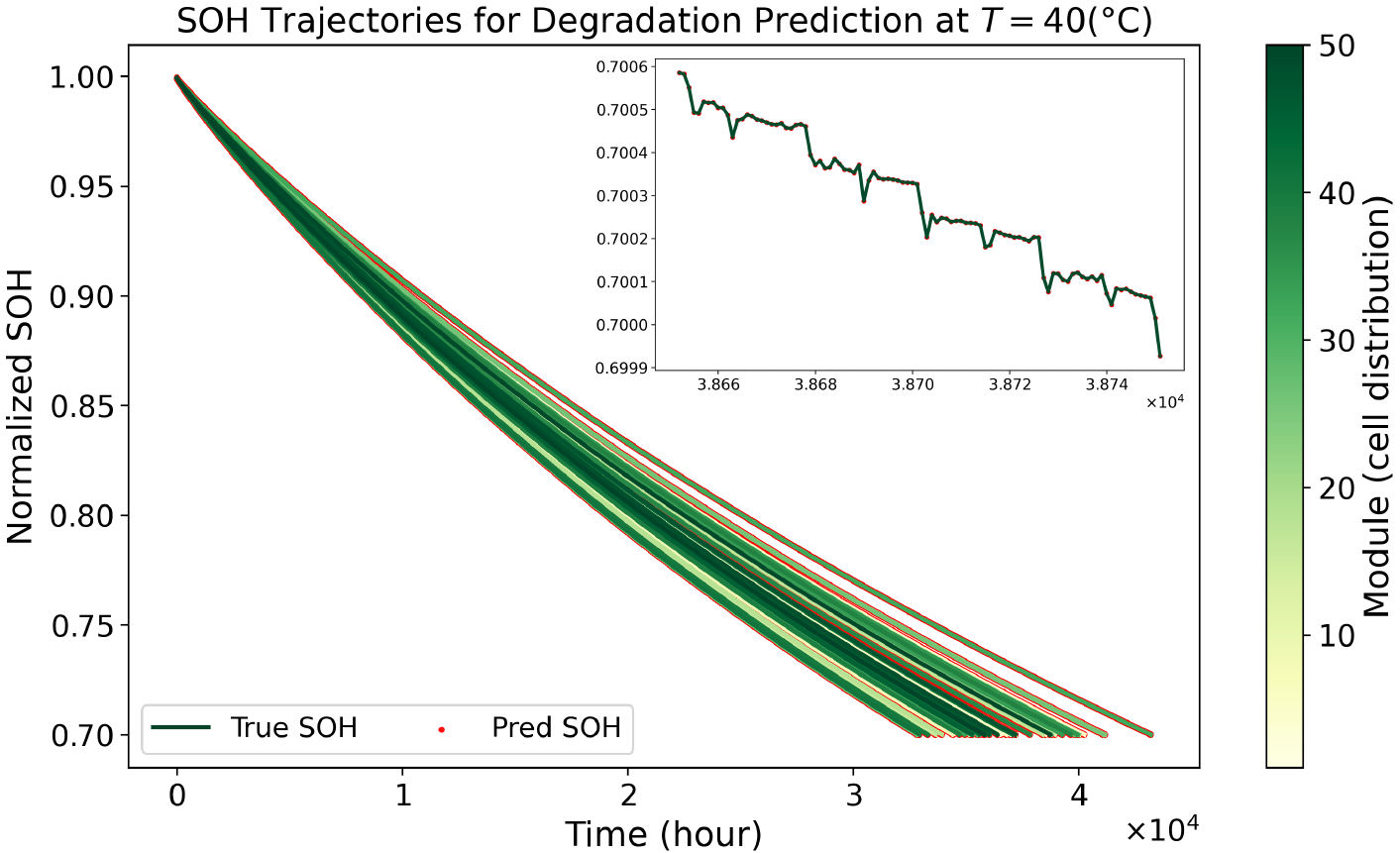}
\put(0,59){(a)}
\end{overpic}\\
\vspace{.05in}
\begin{overpic}[width=0.95\columnwidth]{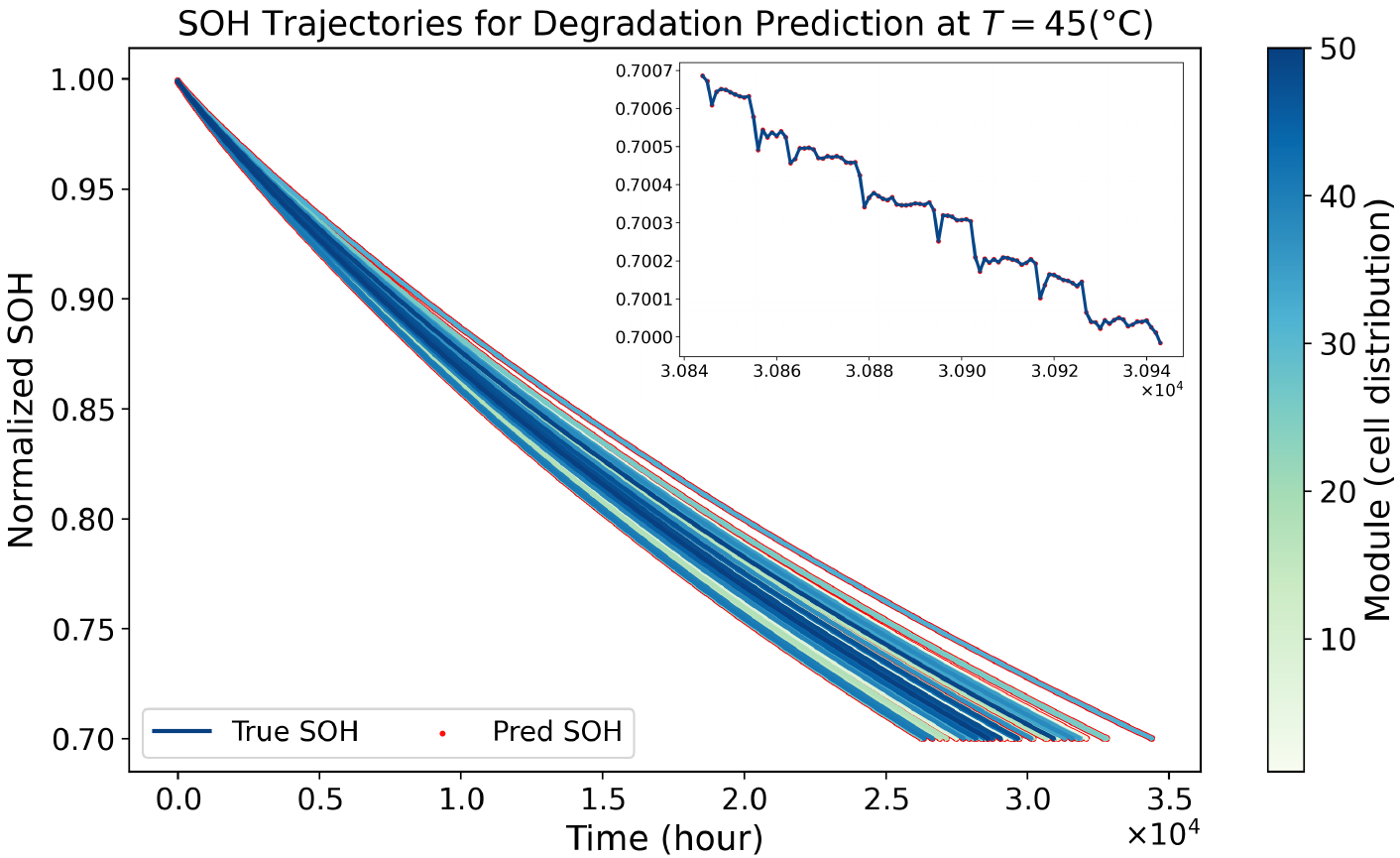}
\put(0,59){(b)}
\end{overpic}\\
\vspace{-.1in}
\caption{Predicted and true SOH trajectories of a battery module operated at $40^{\circ}\mathrm{C}$ and $45^{\circ}\mathrm{C}$. Strong agreement is maintained despite the presence of additive measurement noise, as highlighted in the zoomed-in views.}
\label{fig2}
\vspace{-.175in}
\end{figure}

Figure~\ref{fig2} compares the predicted and reference SOH trajectories for all 50 cells over a three-year degradation period. \emph{StateFormer} accurately captures the long-term degradation behavior across the entire fleet, with the predicted trajectories closely matching the ground truth despite the unseen temperature conditions.
Quantitative comparisons are summarized in Table~\ref{tab1}, which reports the mean and standard deviation of the predicted SOC at the beginning of life ($\mathrm{SOC_{BOL}}$), SOC at the end of life ($\mathrm{SOC_{EOL}}$), and SOH at the end of life ($\mathrm{SOH_{EOL}}$) under $1\%$ additive Gaussian noise. The predicted statistics closely agree with the reference values, with a maximum discrepancy of approximately $0.02\%$ in $\mathrm{SOC_{BOL}}$ and $\mathrm{SOC_{EOL}}$, while the associated standard deviations remain below $0.001\%$, demonstrating the high fidelity and consistency of the forecasting model in learning the Arrhenius behavior.

The robustness of \emph{StateFormer} under progressively noisier measurements is further evaluated in Table~\ref{tab2}, which reports the relative $L_2$ errors and their corresponding $95\%$ confidence intervals for noise levels ranging from $1\%$ to $5\%$. Across all noise levels, \emph{StateFormer} consistently maintains a relative $L_2$ error on the order of $2\times10^{-2}\%$, while the $95\%$ confidence intervals remain below $1\times10^{-4}\%$. These results demonstrate that the proposed framework is highly robust to measurement noise and delivers stable, accurate predictions under realistic sensing uncertainties.

\begin{table}[tbp]
\caption{Statistical comparison of true and predicted BOL/EOL values of SOC and SOH at $45^{\circ}\mathrm{C}$ with $1\%$ additive Gaussian noise.}
\begin{center}
\begin{tabular}{|c|cc|cc|c|}
\hline
\textbf{Variable}&\multicolumn{5}{c|}{\textbf{Battery State of Charge 
\& Health over Time}} \\
\cline{2-6} 
\textbf{Statistics} &
\multicolumn{2}{c|}{$\mathbf{SOC}_{\mathbf{BOL}}\,(\%)$} &
\multicolumn{2}{c|}{$\mathbf{SOC}_{\mathbf{EOL}}\,(\%)$} &
$\mathbf{SOH}_{\mathbf{EOL}}\,(\%)$ \\
&
$\mathbf{Max}$$^{\mathrm{a}}$ &
$\mathbf{Min}$ &
$\mathbf{Max}$ &
$\mathbf{Min}$ &
\\
\hline
$\mu_{True}$& $94.944$ & $6.442$ & $80.912$ & $19.089$ & $69.997$ \\
\hline
$\sigma_{True}$& $0.004$ & $0.004$ & $0.006$ & $0.006$ & $69.997$ \\
\hline
$\mu_{pred}$& $94.961$ & $6.442$ & $80.919$ & $19.072$ &  $0.002$\\
\hline
$\sigma_{pred}$&  $0.005$ & $0.004$ & $0.006$ &  $0.006$ & $0.002$ \\
\hline
\multicolumn{6}{l}{$^{\mathrm{a}}$ Under baseline parameters, the allowable SOC window narrows~\eqref{eqSOCmax}.}
\end{tabular}
\end{center}
\label{tab1}
\vspace{-.2in}
\end{table}

\begin{table}[tbp]
\caption{Relative $L_2$ error and $95\%$ confidence intervals under additive Gaussian noise levels of 1\%,3\%,5\%and 10\%.}
\label{tab2}
\centering
\begin{tabular}{|c|c|c|}
\hline
\textbf{Noise} &
\multicolumn{2}{c|}{\textbf{StateFormer Prediction Performance}} \\
\cline{2-3}
\textbf{Level} &
\textbf{Relative $L_2$ Error (\%)} &
\textbf{95\% CI (\%)} \\
\hline
$\eta = 1\%$ & $1.64e\text{-}02 \pm 1.48e\text{-}04$  &  $4.11e\text{-}05$ \\
\hline
$\eta = 3\%$  & $1.93e\text{-}02 \pm 1.03e\text{-}04$  & $2.84e\text{-}05$  \\
\hline
$\eta = 5\%$  & $2.78e\text{-}02 \pm 3.23e\text{-}04$ & $8.95e\text{-}05$  \\
\hline
$\eta = 10\%$  & $2.81e\text{-}02 \pm 1.42e\text{-}03$ & $3.92e\text{-}04$  \\
\hline
\end{tabular}
\vspace{-.15in}
\end{table}


\subsection{Field measurements of home storage systems}
We further evaluate the predictive performance of \emph{StateFormer} using real-world field measurements from a residential home storage system (HSS) integrated with a rooftop photovoltaic (PV) system~\cite{figgener2024multi}. The available measurements include battery current, voltage, power, battery pack housing temperature, and ambient room temperature. In this study, \emph{StateFormer} is trained to predict long-horizon voltage and battery temperature trajectories from historical current measurements and room temperature.

The dataset spans five years of operation. Measurements collected over the first $35$ months ($2015$–$2018$) are used for training, the subsequent seven months in $2019$ are reserved for validation, and the remaining ten months of data in $2020$ (excluding October and November, for which measurements are unavailable) are used for testing. This chronological partition evaluates the model's ability to generalize to future operating conditions over extended time horizons.

\begin{figure}[tbp]
\centering
\begin{overpic}[width=0.995\columnwidth]{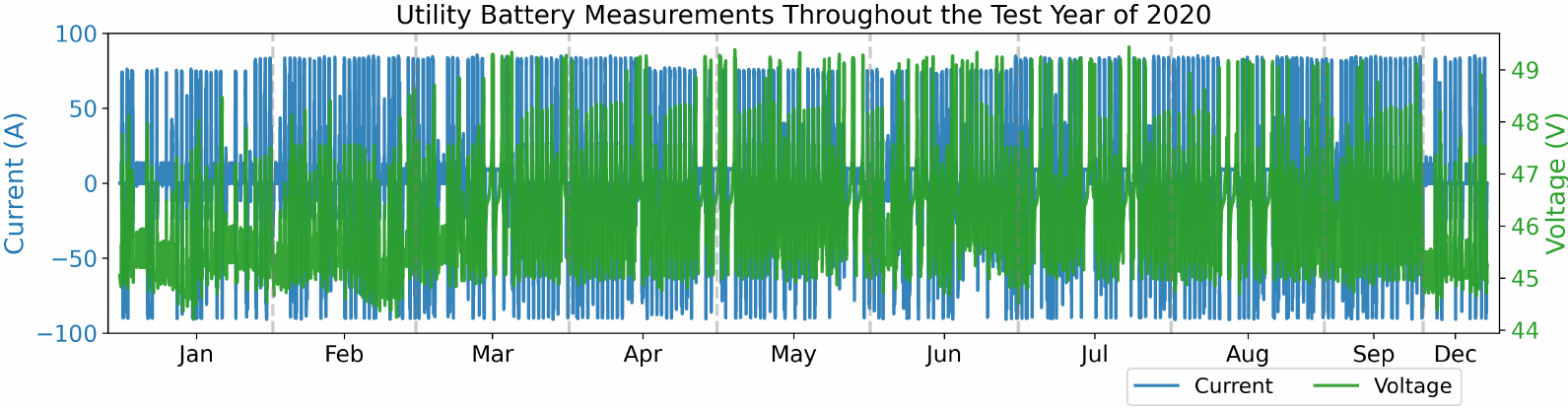}
\put(-1.4,23){\small(a)}
\end{overpic}
\\
\vspace{.05in}
\begin{overpic}[width=0.98\columnwidth]{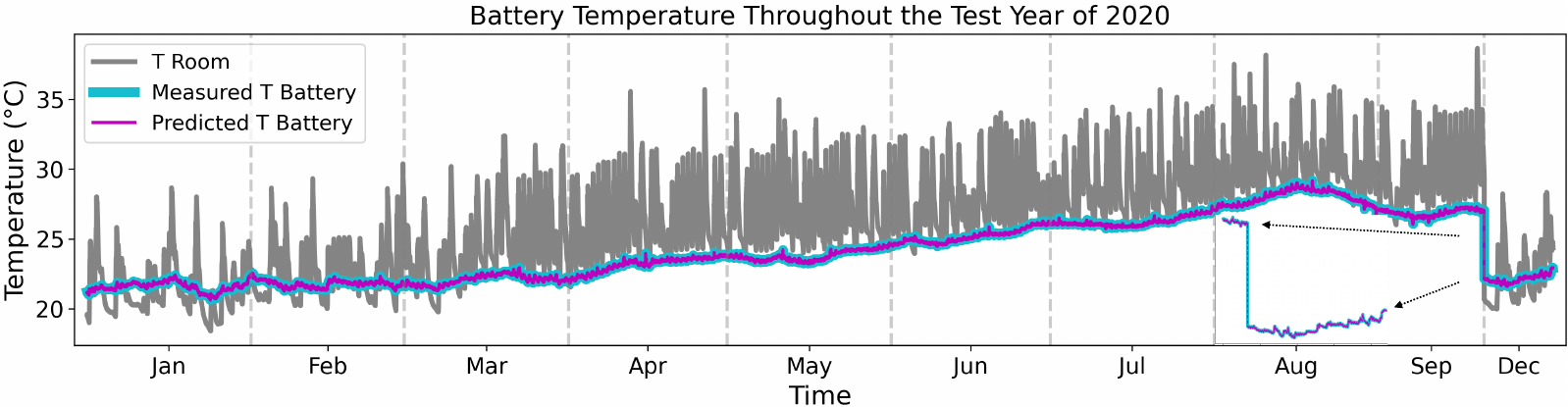}
\put(-1.9,23){\small(b)}
\end{overpic}\\
\vspace{.07in}
\begin{overpic}[width=0.98\columnwidth]{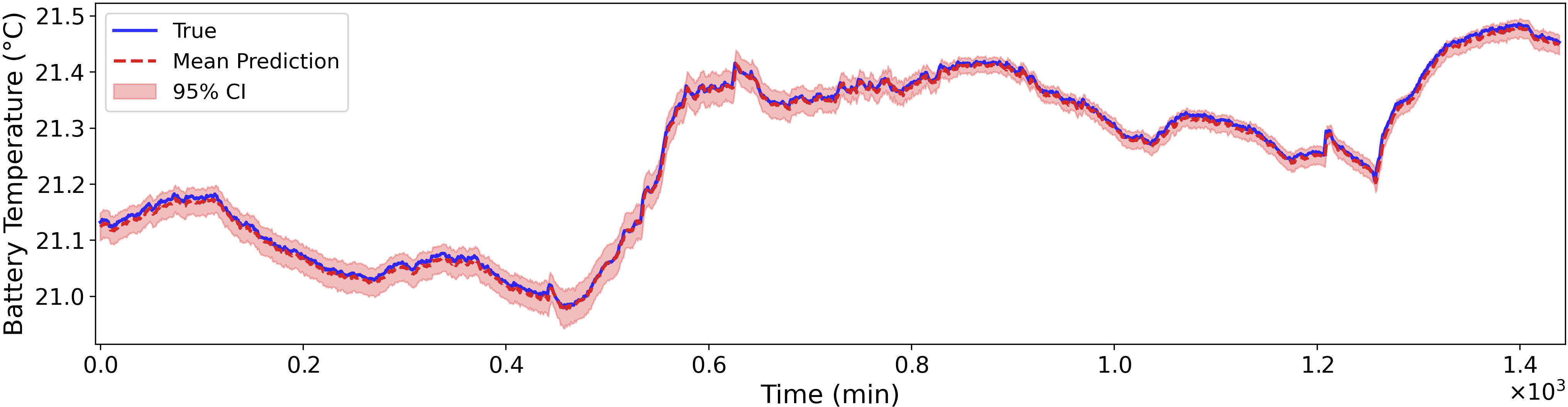}
\put(-1.8,26.5){\small(c)}
\end{overpic}\\
\vspace{.03in}
\begin{overpic}[width=0.98\columnwidth]{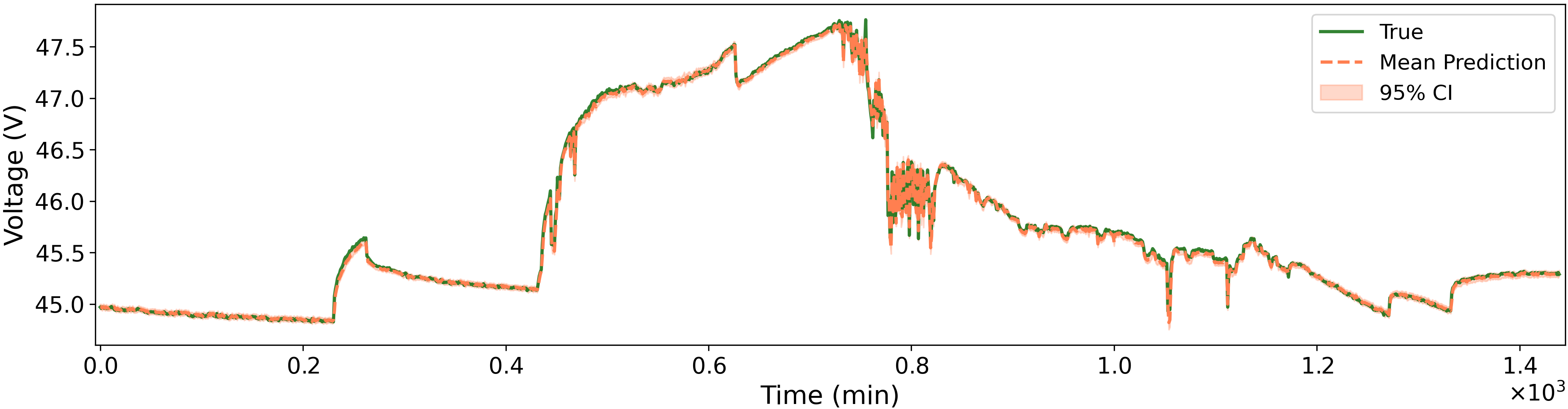}
\put(-1.1,26){\small(d)}
\end{overpic}
\vspace{-.05in}
\caption{Field measurements from the HSS and corresponding \emph{StateFormer} predictions. (a) Measured battery current and voltage over the 2020 operating period. (b) Measured and predicted battery cell temperature using the measured current and room temperature as inputs over the same period. (c-d) Measured battery temperature and voltage during a representative 24-hour test period, showing that the field measurements remain within the predicted uncertainty bands.}
\label{fig3}
\vspace{-.2in}
\end{figure}

Fig.~\ref{fig3} compares the predicted and measured battery temperature and voltage given current and room temperature throughout the $2020$ test period. \emph{StateFormer} accurately reproduces the long-term temporal evolution of both voltage and battery temperature, including the abrupt operating transition observed between September and December. Across five independent training realizations, the model achieves an average relative $L_2$ error of approximately $0.05\%$, with a narrow $95\%$ confidence interval of $0.06\%$. As illustrated in the bottom panel of Fig.~\ref{fig3}, the measured voltage and battery temperature consistently remain within the predicted uncertainty bands over the entire test period, demonstrating both the accuracy and reliability of the proposed framework for long-term battery state forecasting under real operating conditions.

\section{Conclusion}
The proposed Transformer-based autoregressive framework enables accurate, robust, and uncertainty-aware long-horizon battery state forecasting across both synthetic battery fleets and multi-year field measurements, demonstrating strong generalization to unseen operating conditions and measurement noise. By combining historical battery states with latest operating conditions through self-attention, the model captures the nonlinear electrothermal dynamics governing battery degradation while remaining robust to measurement noise. The close agreement between predicted and measured trajectories validates the model's ability to perform accurate long-horizon forecasting while preserving short-term prediction accuracy across diverse operating conditions. These results demonstrate the value of combining synthetic fleet-scale and field datasets to develop data-driven battery digital twins for accurate state estimation, health monitoring, and degradation-aware control and optimization in next-generation electric vehicle and grid-scale energy storage systems.

Beyond forward prediction, \emph{StateFormer} serves as a computationally efficient surrogate for inverse degradation modeling, enabling the rapid synthesis of current input profiles that achieve desired SOH trajectories and target battery lifetimes. This capability provides a foundation for scalable battery digital twins that support real-time fleet health monitoring, predictive diagnostics, and degradation-aware operational optimization. As battery energy storage systems become critical to future power grids and hyperscale data center infrastructures, battery management can no longer rely on static, single-asset approximations. Instead, next-generation energy management systems require AI frameworks that capture coupled electrochemical-thermal dynamics, fleet-level heterogeneity, and long-term degradation to ensure the reliability, resilience, and economic efficiency of large-scale energy storage throughout its service lifetime.

\section*{Acknowledgment}

This research used resources of the National Energy Research Scientific Computing Center (NERSC), a Department of Energy User Facility. We would like to thank Paul Gasper for valuable discussions on the source data. 

\bibliographystyle{IEEEtran}
\bibliography{refs}

@inproceedings{bai2026predicting,
  title={Predicting Wave Reflection and Transmission in Heterogeneous Media via Fourier Operator-Based Transformer Modeling},
  author={Bai, Zhe and Johansen, Hans},
  booktitle={2026 International Conference on Artificial Intelligence, Computer, Data Sciences and Applications (ACDSA)},
  pages={1--6},
  year={2026},
  organization={IEEE}
}

@article{vaswani2017attention,
  title={Attention is all you need},
  author={Vaswani, Ashish and Shazeer, Noam and Parmar, Niki and Uszkoreit, Jakob and Jones, Llion and Gomez, Aidan N and Kaiser, {\L}ukasz and Polosukhin, Illia},
  journal={Advances in neural information processing systems},
  volume={30},
  year={2017}
}

@article{figgener2024multi,
  title={Multi-year field measurements of home storage systems and their use in capacity estimation},
  author={Figgener, Jan and Van Ouwerkerk, Jonas and Haberschusz, David and Bors, Jakob and Woerner, Philipp and Mennekes, Marc and Hildenbrand, Felix and Hecht, Christopher and Kairies, Kai-Philipp and Wessels, Oliver and others},
  journal={Nature Energy},
  volume={9},
  number={11},
  pages={1438--1447},
  year={2024},
  publisher={Nature Publishing Group UK London}
}

@inproceedings{zhou2023deep,
  title={Deep latent state space models for time-series generation},
  author={Zhou, Linqi and Poli, Michael and Xu, Winnie and Massaroli, Stefano and Ermon, Stefano},
  booktitle={International Conference on Machine Learning},
  pages={42625--42643},
  year={2023},
  organization={PMLR}
}

@article{bai2020dynamic,
  title={Dynamic mode decomposition for compressive system identification},
  author={Bai, Zhe and Kaiser, Eurika and Proctor, Joshua L and Kutz, J Nathan and Brunton, Steven L},
  journal={AIAA Journal},
  volume={58},
  number={2},
  pages={561--574},
  year={2020},
  publisher={American Institute of Aeronautics and Astronautics}
}

@inproceedings{kumtepeli2024depreciation,
  title={Depreciation cost is a poor proxy for revenue lost to aging in grid storage optimization},
  author={Kumtepeli, Volkan and Hesse, Holger and Morstyn, Thomas and Nosratabadi, Seyyed Mostafa and Aunedi, Marko and Howey, David A},
  booktitle={2024 American Control Conference (ACC)},
  pages={701--706},
  year={2024},
  organization={IEEE}
}

@inproceedings{koller2013defining,
  title={Defining a degradation cost function for optimal control of a battery energy storage system},
  author={Koller, Michael and Borsche, Theodor and Ulbig, Andreas and Andersson, G{\"o}ran},
  booktitle={2013 IEEE Grenoble Conference},
  pages={1--6},
  year={2013},
  organization={IEEE}
}

@article{rahman2024exploring,
  title={Exploring lithium-ion battery degradation: A concise review of critical factors, impacts, data-driven degradation estimation techniques, and sustainable directions for energy storage systems},
  author={Rahman, Tuhibur and Alharbi, Talal},
  journal={Batteries},
  volume={10},
  number={7},
  pages={220},
  year={2024},
  publisher={MDPI}
}

@article{fasolato2025analyzing,
  title={Analyzing cell-to-cell heterogeneities and cell configurations in parallel-connected battery modules using physics-based modeling},
  author={Fasolato, Simone and Allam, Anirudh and Onori, Simona and Raimondo, Davide M},
  journal={Journal of Energy Storage},
  volume={129},
  pages={116942},
  year={2025},
  publisher={Elsevier}
}

@inproceedings{shete2021battery,
  title={Battery management system for soc estimation of lithium-ion battery in electric vehicle: a review},
  author={Shete, Suwarna and Jog, Pranjal and Kumawat, RK and Palwalia, DK},
  booktitle={2021 6th IEEE International Conference on Recent Advances and Innovations in Engineering (ICRAIE)},
  volume={6},
  pages={1--4},
  year={2021},
  organization={IEEE}
}

@article{che2025diagnostic,
  title={Diagnostic-free onboard battery health assessment},
  author={Che, Yunhong and Lam, Vivek N and Rhyu, Jinwook and Schaeffer, Joachim and Kim, Minsu and Bazant, Martin Z and Chueh, William C and Braatz, Richard D},
  journal={Joule},
  volume={9},
  number={8},
  year={2025},
  publisher={Elsevier}
}

@article{liu2025warranties,
  title={Warranties of batteries: Requirements, state-of-the-art, relevant analysis methods, and research perspectives},
  author={Liu, Yiliu and Wang, Xiao-Lin},
  journal={Journal of Reliability Science and Engineering},
  volume={1},
  number={3},
  pages={032003},
  year={2025},
  publisher={IOP Publishing}
}

@article{how2019state,
  title={State of charge estimation for lithium-ion batteries using model-based and data-driven methods: A review},
  author={How, Dickson NT and Hannan, MA and Lipu, MS Hossain and Ker, Pin Jern},
  journal={Ieee Access},
  volume={7},
  pages={136116--136136},
  year={2019},
  publisher={IEEE}
}

@article{wang2021review,
  title={A review on online state of charge and state of health estimation for lithium-ion batteries in electric vehicles},
  author={Wang, Zuolu and Feng, Guojin and Zhen, Dong and Gu, Fengshou and Ball, Andrew},
  journal={Energy Reports},
  volume={7},
  pages={5141--5161},
  year={2021},
  publisher={Elsevier}
}

@article{severson2019data,
  title={Data-driven prediction of battery cycle life before capacity degradation},
  author={Severson, Kristen A and Attia, Peter M and Jin, Norman and Perkins, Nicholas and Jiang, Benben and Yang, Zi and Chen, Michael H and Aykol, Muratahan and Herring, Patrick K and Fraggedakis, Dimitrios and others},
  journal={Nature energy},
  volume={4},
  number={5},
  pages={383--391},
  year={2019},
  publisher={Nature Publishing Group UK London}
}


\end{document}